\documentclass[aps,twocolumn,showpacs,superscriptaddress,amsmath,amssymb,amsfonts,floatfix,longbibliography]{revtex4-2}

\usepackage[T1]{fontenc} 
\usepackage{graphicx}
\usepackage{float}
\usepackage{dcolumn}
\usepackage{bm}
\usepackage{amssymb}
\usepackage{tabularx}
\usepackage{microtype}
\usepackage{xfrac}
\usepackage{chemformula}
\usepackage{gensymb}
\usepackage[most]{tcolorbox}
\usepackage{xcolor}
\usepackage{multirow}
\usepackage{enumitem}
\usepackage{array}
\usepackage{booktabs}
\usepackage[flushleft]{threeparttable}
\usepackage{natbib,hyperref}
\usepackage{graphicx}
\usepackage{makecell}
\usepackage{colortbl}
\usepackage{tcolorbox} 
\usepackage{xcolor}
\usepackage{soul}

\newcommand{\angstrom}{\mbox{\normalfont\AA}}

\hypersetup{
	colorlinks=true, 
	citecolor=blue,  
}

\begin{document}

\title{Coherent high-velocity chiral magnons in the metallic altermagnet CrSb}

\author{Ashutosh K. Singh}
    \affiliation{Quantum Matter Institute, University of British Columbia, Vancouver, BC V6T 1Z4, Canada}
    \affiliation{Department of Physics \& Astronomy, University of British Columbia, Vancouver, BC V6T 1Z1, Canada}

\author{Niclas Heinsdorf}
    \affiliation{Quantum Matter Institute, University of British Columbia, Vancouver, BC V6T 1Z4, Canada}
    \affiliation{Department of Physics \& Astronomy, University of British Columbia, Vancouver, BC V6T 1Z1, Canada}
    \affiliation{Max Planck Institute for Solid State Research, Heisenbergstrasse 1, 70569 Stuttgart, Germany}

\author{Abraham A. Mancilla}
    \affiliation{Quantum Matter Institute, University of British Columbia, Vancouver, BC V6T 1Z4, Canada}
    \affiliation{Department of Physics \& Astronomy, University of British Columbia, Vancouver, BC V6T 1Z1, Canada}
    
\author{Jörn Bannies}
    \affiliation{Quantum Matter Institute, University of British Columbia, Vancouver, BC V6T 1Z4, Canada}
    \affiliation{Department of Chemistry, University of British Columbia, Vancouver, BC V6T 1Z1, Canada}
    
\author{Avishek Maity}
    \affiliation{Neutron Scattering Division, Oak Ridge National Laboratory, Oak Ridge, TN 37831, USA}
    
\author{Alexander I. Kolesnikov}
    \affiliation{Neutron Scattering Division, Oak Ridge National Laboratory, Oak Ridge, TN 37831, USA}

\author{Masaaki Matsuda}
    \affiliation{Neutron Scattering Division, Oak Ridge National Laboratory, Oak Ridge, TN 37831, USA}
    
\author{ Matthew B. Stone}
    \affiliation{Neutron Scattering Division, Oak Ridge National Laboratory, Oak Ridge, TN 37831, USA}    
    
\author{Marcel Franz}
    \affiliation{Quantum Matter Institute, University of British Columbia, Vancouver, BC V6T 1Z4, Canada}
    \affiliation{Department of Physics \& Astronomy, University of British Columbia, Vancouver, BC V6T 1Z1, Canada}
    \affiliation{Canadian Institute for Advanced Research (CIFAR), Toronto, ON, M5G 1M1, Canada}

\author{Jonathan Gaudet}
    \affiliation{NIST Center for Neutron Research, National Institute of Standards and Technology, Gaithersburg, MD 20899, USA}
    \affiliation{Department of Materials Science and Eng., University of Maryland, College Park, MD 20742-2115}

\author{Alannah M. Hallas}
\email[Email: ]{alannah.hallas@ubc.ca}
    \affiliation{Quantum Matter Institute, University of British Columbia, Vancouver, BC V6T 1Z4, Canada}
    \affiliation{Department of Physics \& Astronomy, University of British Columbia, Vancouver, BC V6T 1Z1, Canada}
    \affiliation{Canadian Institute for Advanced Research (CIFAR), Toronto, ON, M5G 1M1, Canada}

\date{\today}
\begin{abstract}

Altermagnets combine the zero net magnetization of antiferromagnets with momentum-dependent spin splitting, creating new opportunities for spintronic and magnonic technologies. The full realization of this promise necessitates a material that is simultaneously metallic and altermagnetic, yet preserves the long magnon lifetimes more often seen in insulators. Here, we use polarized neutron diffraction and inelastic neutron scattering to determine the magnetic structure and spin dynamics of the metallic altermagnet CrSb. We confirm a fully compensated, collinear magnetic structure with localized Cr moments and establish the microscopic spin Hamiltonian governing its excitations. Despite its metallicity, CrSb hosts remarkably coherent magnons across its 230 meV bandwidth, with group velocities approaching 60 km/s. Along low-symmetry momentum directions, we resolve chiral magnon splitting generated by higher-order, symmetry-inequivalent exchange interactions, providing the first experimental observation of this phenomenon in a metallic altermagnet. These results reveal a nearly complete separation between magnetic and electronic degrees of freedom in CrSb and demonstrate that metallic transport, room-temperature altermagnetic order, fast coherent magnons, and chiral spin dynamics can coexist in a single material, positioning CrSb as a compelling platform for spintronic applications.
 
\end{abstract}

\maketitle

Altermagnets are a class of materials that possess a unique combination of magnetic and spatial symmetries~\cite{vsmejkal2022beyond}. Despite having zero net magnetic moment due to their fully compensated antiferromagnetic (AFM) order, the hallmark feature of an altermagnet is non-relativistic spin splitting (NRSS) in the electronic bands~\cite{ahn2019antiferromagnetism,hayami2019momentum,vsmejkal2020crystal,hayami2020bottom,ma2021multifunctional,naka2021perovskite}, a characteristic typically associated with ferromagnets (FMs). Unlike conventional AFMs, the “up” and “down” sites in an altermagnet are not related by fractional translation or inversion symmetries that would otherwise enforce global spin degeneracy. Instead, the AFM order is stabilized by rotational symmetries and spin degeneracy persists only at momenta invariant under these rotations, giving rise to the eponymous alternating pattern of spin chirality in reciprocal space and magnetization density in real space.

\begin{figure*}[htbp]
\centering
\includegraphics[width=0.95\textwidth]{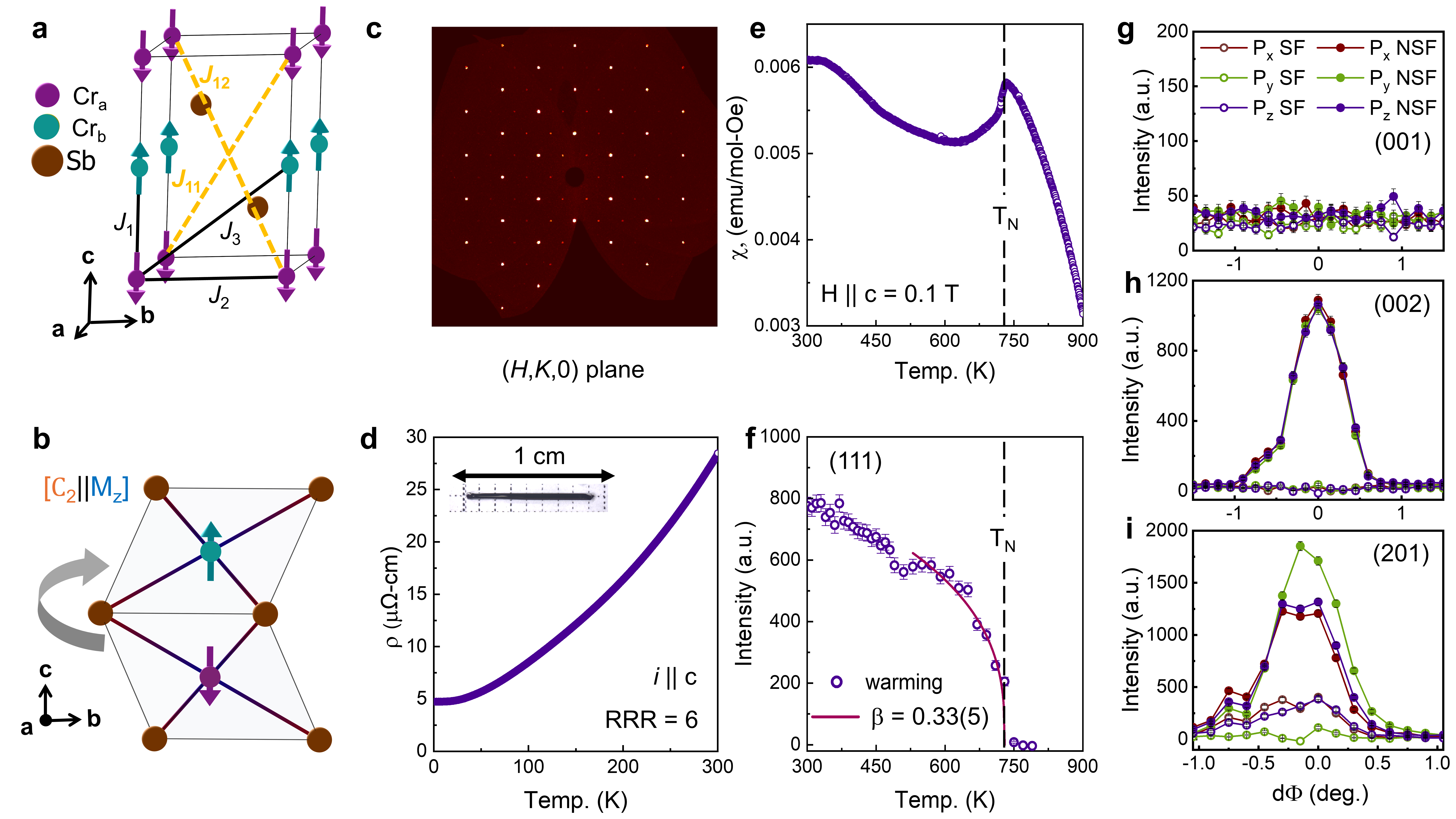}
\caption{\textbf{Altermagnetic order in CrSb.} \textbf{(a)} Crystal structure and altermagnetic ordered state of CrSb where the Cr spins are parallel to the \textit{c}-axis. The nearest-neighbor exchange pathways $J_1$ (AFM), $J_2$ (FM), and $J_3$ (AFM) are shown as solid black lines, while the dashed yellow line represents the weaker $J_{11}$ and $J_{12}$ exchange paths responsible for the altermagnetic splitting in the magnon spectrum. \textbf{(b)} The two opposite-spin sublattices, Cr$_{\mathrm{a}}$ and Cr$_{\mathrm{b}}$, are connected by a $C_2$ spin-space and $M_z$ real-space mirror symmetry. \textbf{(c)} Room temperature ($H$,$K$,0) precession image from single crystal x-ray diffraction. \textbf{(d)} Longitudinal resistivity measured with $i\parallel c$ showing metallic behavior with RRR = 6 (inset: optical image of a needle-shaped crystal of length close to 1 cm). \textbf{(e)} High-temperature magnetic susceptibility measured with $H\parallel c = 0.10$ T, showing the onset of AFM order at $T_N = 733(4)$ K. \textbf{(f)} Temperature-dependent intensity of the (111) magnetic Bragg peak obtained with polarized neutron diffraction.
Polarization dependence of the \textbf{(g)} (001), \textbf{(h)} (002), and \textbf{(i)} (201) Bragg peaks collected at $T = 25$ K, confirming the fully compensated altermagnetic ordered state shown in panel \textbf{(a)}.}
\label{fig:Structure}
\end{figure*}

Altermagnets are the ideal materials platform to enable future spintronic device applications~\cite{vsmejkal2022emerging}. In this context, metallic altermagnets are especially sought-after because they can be used for ultrafast magneto-resistive switching~\cite{vsmejkal2022giant}. Moreover, the NRSS of Fermi surfaces in altermagnetic metals can generate a range of exotic phenomena, including spin-current  generation~\cite{naka2019spin,ma2021multifunctional,naka2021perovskite,shao2021spin,sourounis2025efficient,bf1n-sxdl,52wh-1z5y,rhmg-j1fv} and non-linear responses~\cite{fang2024quantum,ezawa2024intrinsic,mcclarty2024landau,heinsdorf2025altermagnetic}. {\color{black} Altermagnetic metals are therefore ideal candidates for high-frequency magnonic devices, where magnon polarization and direction-dependent lifetimes could enable nonreciprocal transport and fast, field-free switching}~\cite{wang2019switching,yu2021magnetic,pirro2021advances,gomonay2024structure,shiota2024handedness,Choi2025,weber2026nonreciprocal}. 

The emergence of the field of altermagnetism immediately prompted a rediscovery of many decades-old materials, whose symmetry properties had previously been overlooked. Within this context, NRSS in the electronic band structure was quickly established in a host of altermagnets~\cite{lee2024broken,krempasky2024altermagnetic,dale2024non,reimers2024direct,yang2025three,jiang2025metallic,zhang2024crystal}. However, observations of chiral magnon bands have lagged behind and there are only a handful of materials in which spin-split magnons have been reported, such as local moment insulators MnTe ~\cite{liu2024chiral,doi:10.1126/science.adv3196,m8lc-f8gk}, $\alpha$-Fe$_2$O$_3$ \cite{7yhz-jptc} and MnF$_2$~\cite{doi:10.1126/science.aec2602}. To date, experimental signatures of chiral magnons in metallic altermagnets have remained elusive, with their detection complicated by coupling between magnetic and electronic degrees of freedom, which serve to limit magnon lifetimes.

In this paper, we report the first example of NRSS of magnon bands in a metallic altermagnet, CrSb. In addition to its metallicity, CrSb has earth-abundant constituent elements, magnetically orders well-above room temperature, and is air-stable, making it eminently suitable for the above-mentioned classes of applications. Using polarized neutron diffraction, we establish that the AFM order in CrSb is fully localized, with negligible coupling to the conduction electrons. As a consequence, the spin excitation spectrum is highly coherent over its entire bandwidth with gigantic magnon velocities, ideal for ultra-fast information transport.  Finally, through careful analysis of the low-symmetry paths of its spin wave spectrum, we establish the presence of chiral spin-split magnons, a distinctive characteristic of altermagnetism.

\begin{figure*}[htbp]
\centering
\includegraphics[width=0.95\textwidth]{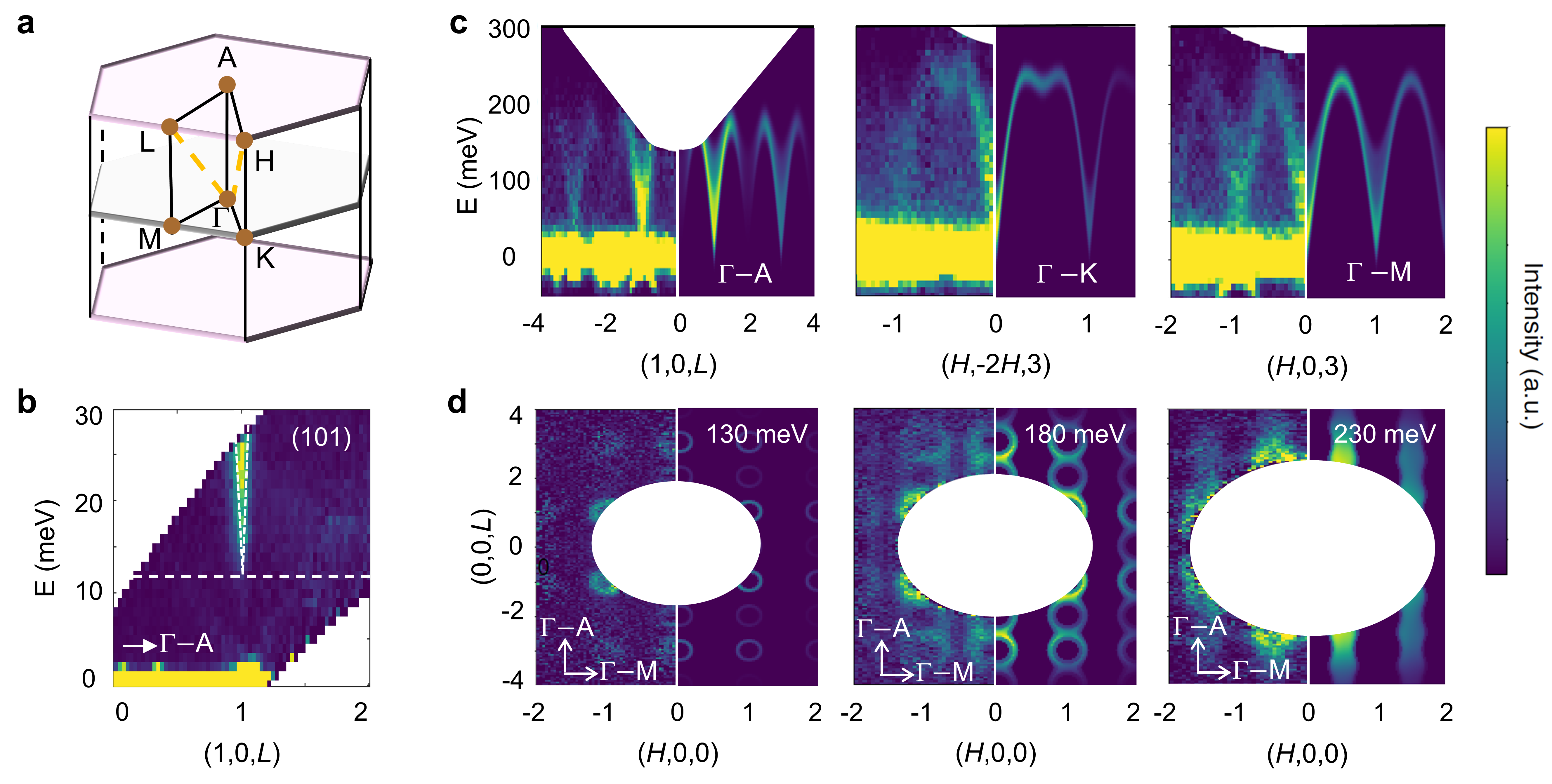}
\caption{\textbf{Determination of the spin Hamiltonian of CrSb from inelastic neutron scattering.}
\textbf{(a)} Brillouin zone of CrSb with indicated high-symmetry directions.
\textbf{(b)} INS spectrum of CrSb measured at 5~K around the (101) Bragg position with an incident neutron energy $E_i$ = 50 meV, revealing a magnon gap of 13(2) meV. The conical dashed lines represent the low-energy spin excitations simulated using the CrSb spin Hamiltonian described in the text.  
\textbf{(c)} The left panels are INS spectra of CrSb collected with $E_i$ = 750 meV along three high-symmetry directions. The right panels show the corresponding simulated spectra based on the fitted spin Hamiltonian. \textbf{(d)} The left panels show constant-energy slices within the ($H,0,L$) plane at 130, 180, and 230 meV. The right panels are calculated slices using the fitted CrSb spin Hamiltonian for comparison.}
\label{fig:Magnon}
\end{figure*}

CrSb crystallizes in the hexagonal space group $P6_3/mmc$, with two distinct magnetic sublattices (defined by Cr atoms with antiparallel spins) labeled as Cr$_a$ and Cr$_b$, located at $\mathbf{r}=(0,0,0)$ and $\mathbf{r}=(0,0,\sfrac{1}{2})$ (Fig.~\ref{fig:Structure}(a)). The local coordination polyhedra surrounding the Cr$_a$ and Cr$_b$ sites can be superimposed by the application of $C_2$ spin-space 
and $M_z$ real-space mirror symmetry operations (Fig.~\ref{fig:Structure}(b)). The high crystalline quality of our CrSb single crystals was confirmed through the refinement of single crystal x-ray diffraction data (Fig.~\ref{fig:Structure}(c), Table SI and SII). Crystal morphology was controlled by the choice of transport reagent, with CrCl$_3$ yielding needle-shaped crystals ideal for transport studies (inset of Fig.~\ref{fig:Structure}(d)) and I$_2$  yielding massive plate-like crystals suitable for diffraction (Fig. S1(a)). Detailed experimental methods are provided in the Supplemental Material~\footnote{See Supplemental Material at [URL to be inserted by publisher] for additional methodological and sample characterization details in Section I. This includes details of the CrSb crystal growth method, structural characterization with powder and single crystal x-ray diffraction, compositional analysis with energy dispersive x-ray spectroscopy, electrical resistivity, magnetometry, polarized neutron diffraction, and inelastic neutron scattering. Section II  provides additional details on the analysis and modeling of the inelastic neutron scattering data, including the parameterization of the microscopic Hamiltonian and the determination of the resolution function. See also references~\cite{RODRIGUEZCARVAJAL199355,Gelato:a27733,Hubschle:kk5092,PhysRevB.111.144402,10.1063/5.0240434,PhysRevB.76.184406,Perring1994MAPS,MARSEGUERRA1959140,ARNOLD2014156} therein.}. \

The temperature dependence of the longitudinal electric resistivity ($\rho$), collected with current $i \parallel c$, and the magnetic susceptibility ($\chi$), measured with applied field $H \parallel c$, are reported in Fig.~\ref{fig:Structure}(d) and (e), respectively. The resistivity is nearly isotropic (Fig. S3) with a residual resistivity ratio (RRR) of 6, consistent with previous reports~\cite{https://doi.org/10.1002/advs.202502226}. The $\chi(T)$ measurement shows a sharp cusp at the N\'eel temperature of $T_N$ = 733(4)~K, marking the transition to an antiferromagnetically ordered state. 

Although the AFM magnetic structure of CrSb has been known since the 1950s~\cite{PhysRev.85.365}, previous magnetic structure determinations employed unpolarized neutron diffraction on polycrystalline samples~\cite{takei1963magnetic,PhysRev.85.365}. Importantly, these studies also predate knowledge of altermagnetism. We therefore performed single-crystal polarized neutron diffraction to distinguish between nuclear and magnetic Bragg scattering and to validate that CrSb does exhibit a perfectly compensated altermagnetic order. These measurements were performed using the HB-1 Polarized Triple-Axis Spectrometer at the High Flux Isotope Reactor at Oak Ridge National Laboratory (ORNL). We collected spin-flip (SF) and non-spin-flip (NSF) scattering cross-sections for all accessible Bragg peaks lying within the $(H,0,L)$ and $(H,H,L)$ scattering planes. For each of these peaks, we acquired cross-sections for three different polarization directions ($\mathbf{P_x}$,$\mathbf{P_y}$,$\mathbf{P_z}$) chosen such that $\mathbf{P_x}\parallel\mathbf{Q}$, $\mathbf{P_z}\parallel$(1$\overline{2}$0) or $\mathbf{P_z}\parallel$(1$\overline{1}$0) depending on the scattering plane, and $\mathbf{P_y}=\mathbf{P_z}\times~\mathbf{P_x}$. 

We first confirmed $T_N$ by collecting a series of temperature-dependent rocking scans centered at $\mathbf{Q}=(111)$. As it is only sensitive to the magnetic scattering, we used the SF $\mathbf{P_x}$ polarization channel for this purpose. The integrated intensity of each rocking scan is plotted as a function of temperature in Fig.~\ref{fig:Structure}(f), and reveals a transition at $T_N$ with a critical exponent $\beta=0.33(5)$. This exponent is distinguishable from the mean-field $\beta=0.5$ value and agrees well with the value expected for the 3D-Ising ($\beta=0.326$) universality class.\

Next, we measured all six polarized neutron scattering cross-sections for 7 distinct Bragg reflections at 25~K, within the magnetically ordered phase of CrSb. For a $\mathbf{k} = 0$ magnetic order, parallel and antiparallel spin arrangements between the two Cr magnetic sublattices give rise to magnetic Bragg intensities at $(H0L)$ reflections with $even$ and $odd$ values of $L$, respectively. As shown in Fig.~\ref{fig:Structure}(g-i) and Fig. S4, we observed SF scattering exclusively at Bragg peaks with $odd$ $L$ values, such as $(101)$, $(201)$, and $(103)$. On the other end, peaks with $even$ $L$ values, such as $(100)$, $(102)$, and $(002)$, only have NSF scattering, with equal intensity across all three polarization channels, implying that they are purely nuclear. These observations demonstrate that Cr spins are fully anti-parallel with no detectable ferromagnetic component.\

Our observations unambiguously point to a $g$-wave altermagnetic ground-state order for CrSb in agreement with the previous reports~\cite{PhysRev.85.365,takei1963magnetic,D0DT03277H}, corresponding to the structure shown in Fig.~\ref{fig:Structure}(a), with all spins pointing exactly along the $c$-axis. This conclusion is supported by two key observations: (1) no detectable magnetic Bragg scattering was observed at $\mathbf{Q}=(00L)$ positions, which would be expected for any in-plane spin components, and (2) magnetic scattering for the NSF was exclusively observed in the $\mathbf{P_y}$ channel, which is expected for out-of-plane spins. In contrast, in-plane spin components would produce stronger NSF scattering in the $\mathbf{P_z}$ channels. The fact that we do not observe any intensity difference between $\mathbf{P_z}$ SF and $\mathbf{P_x}$ SF, thus shows that the spins are pointing purely out-of-plane. We refine an ordered moment of 3.2(4)~$\mu_B$/Cr, confirming that the magnetism in CrSb is highly localized in spite of its metallic nature.

\begin{figure*}[htbp]
\centering
\includegraphics[width=0.95\textwidth]{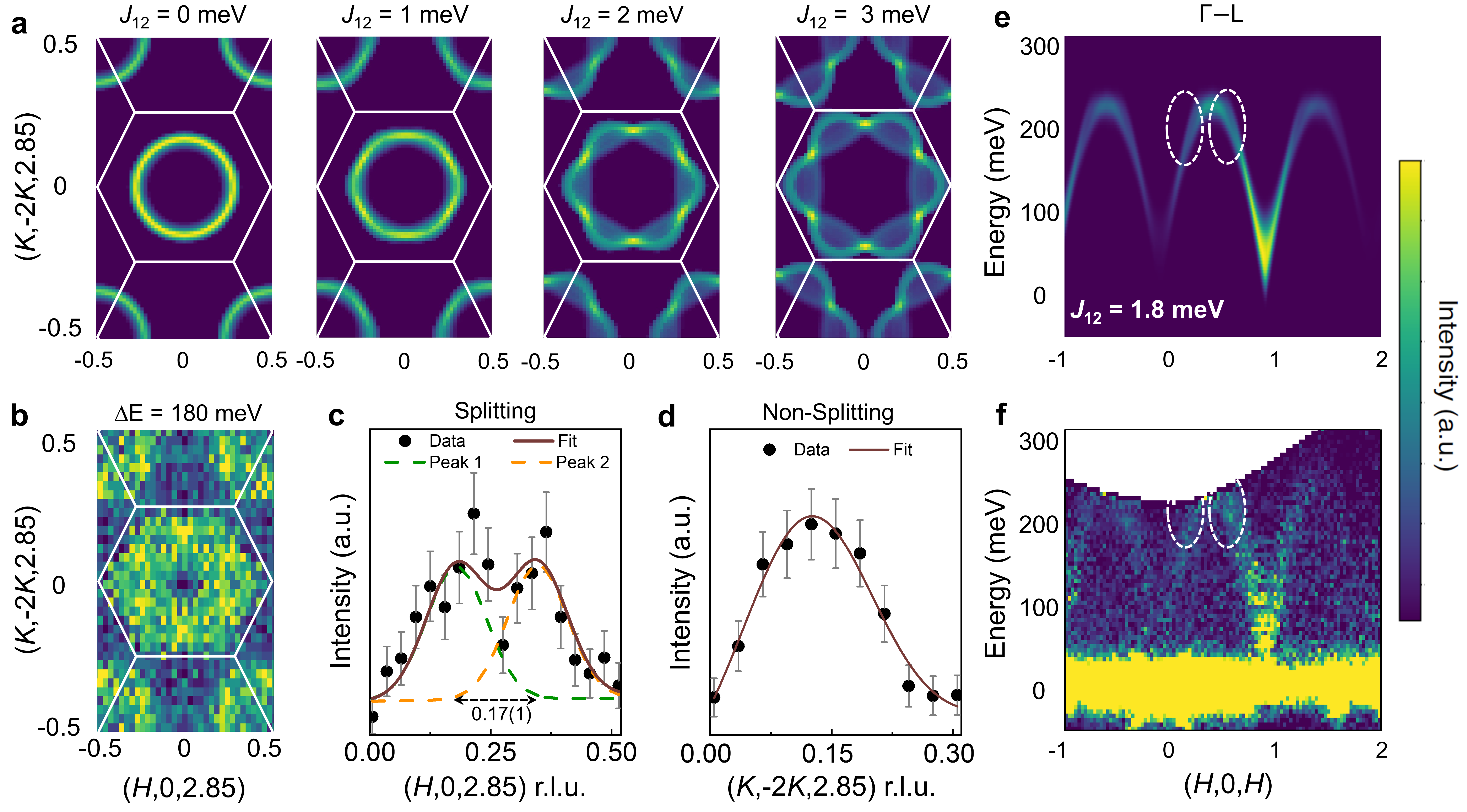}
\caption{\textbf{Observation of spin-split magnons.} 
\textbf{(a)} Simulated constant-energy slices within the ($H$$+$$K,\text{-}2K$,2.85) plane at 180 meV, illustrating the evolution of the spin excitation upon increasing the $J_{12}$ exchange parameter.
\textbf{(b)} Experimental constant-energy slice acquired within the ($H$$+$$K$,$\text{-}2K$,2.85) plane at 180 meV. The white line denotes the Brillouin zone boundary. The 6-fold intensity distribution highlights the presence of splitting along the $(H,0,2.85)$ and its absence along the ($K$,$\text{-}2K$,2.85) direction. 
\textbf{(c,d)} A Gaussian fit to the 180 meV 1D intensity profile along $(H,0,2.85)$ gives the splitting of 0.30(2)~\AA$^{-1}$ (0.17(1) r.l.u.) compared to the instrument resolution of 0.191(1)~\AA$^{-1}$, while the absence of splitting along the ($K$,-2$K$,2.85) direction is a signature of altermagnetism. This observed splitting in $Q$-space can be produced by $J_{12}$ = 1.8 meV.
\textbf{(e)} Simulated INS spectra for the $\Gamma$-L direction, where the white dashed ellipsoid shows the location of the magnon splitting, assuming $J_{12}$ = 1.8 meV. 
\textbf{(f)} Experimental INS data acquired along the $\Gamma$-L direction matches well with simulated data but lacks the energy resolution to clearly resolve the splitting seen in the constant energy slice.}
\label{fig:Splitting}
\end{figure*}

We next characterize the magnetic excitations in CrSb using unpolarized inelastic neutron scattering (INS). Measurements were performed on the SEQUOIA time-of-flight spectrometer at ORNL ~\cite{granroth2010sequoia}, using approximately 45 co-aligned single crystals with a total mass of 5 g. Due to its Ising order, the magnon spectra of CrSb is expected to be gapped, which we probed using neutrons with incident energy $E_i = 50$~meV, focusing on the magnetic zone center $(101)$ (Fig.~\ref{fig:Magnon}(b)). The minimum of a sharply dispersing spin wave mode is gapped from the elastic line by 13(2) meV (Fig. S5), in agreement with previous reports~\cite{PhysRevB.54.11940}.

The full magnon dispersion across the entire Brillouin zone (BZ) of CrSb (Fig.~\ref{fig:Magnon}(a)) is clearly resolved using the $E_i = 750$~meV dataset, shown in Fig.~\ref{fig:Magnon}(c) for three high-symmetry momentum paths corresponding to the $\Gamma$–A, $\Gamma$–K, and $\Gamma$–M directions. The scattered neutron intensity associated with the gapped excitations systematically decreases with increasing momentum transfer, consistent with their magnetic origin. As expected for an AFM, the magnon spectrum exhibits a linear dispersion at low energies. These magnons are observed to be highly coherent across multiple BZs.

The spin excitations in CrSb are anisotropic, with an in-plane bandwidth of approximately 230 meV and an out-of-plane bandwidth of about 180~meV. Constant-energy slices within the ($H,0,L$) scattering plane, which corresponds to the $\Gamma$–M–L–A plane of the BZ, are shown in Fig.~\ref{fig:Magnon}(d) for energy transfers of 130, 180, and 230~meV. The 130 and 180 meV slices reveal well-defined ellipsoids, indicative of long-lived and coherent spin-wave excitations. From the principle axes of the 130~meV ellipsoids, we extract in-plane and out-of-plane magnon group velocities of 408(9)~meV$\cdot\angstrom$ (61(2) km s$^{-1}$) and 386(2)~meV$\cdot\angstrom$ (58(2) km s$^{-1})$, respectively \cite{RevModPhys.60.209}. These measured magnon group velocities exceed those of $\alpha$-Fe$_2$O$_3$ (22.5 km s$^{-1}$) and yttrium-iridium-garnets (13 km s$^{-1}$), which are used in prospective THz and AFM magnonic devices~\cite{PhysRevLett.130.096701,Hortensius2021}. At 230~meV, the scattering is broadly distributed near the BZ boundaries, consistent with the spin-wave dispersion reaching its upper limit.\

The full characterization of the spin wave spectrum of CrSb allows for the determination of its spin Hamiltonian. A minimal model includes near neighbor (n.n.) exchange interactions and single-ion anisotropy given by:
\begin{equation}
\mathcal{H} = -\sum_{\langle i,j \rangle} J_{ij} \, \mathbf{S}_i \cdot \mathbf{S}_j
 - D \sum_{i} (S_i^{c})^{2}
\end{equation}
where $i$ denotes the index of a specific Cr ion with spin operator $\mathbf{S_i}=\sfrac{3}{2}$. $J_{ij}$ is the exchange interactions between the $i_{th}$ and $j_{th}$ Cr spins and the sum is taken over pairs of spins. $D$ is a single-ion anisotropy constant, while $S^c$ is the $c$-axis component of the spin operator. A positive (AFM) $1^{st}$-n.n. $J_1$, a negative (FM) $2^{nd}$-n.n. $J_2$, and a positive $D$ can energetically stabilize the altermagnetic structure of CrSb (Fig.~\ref{fig:Structure}(a)). We found that the inclusion of a positive (AFM) $3^{rd}$-n.n. $J_3$ interaction was required to accurately capture the dispersion of CrSb, which we calculated with linear-spin wave theory using SpinW~\cite{Toth_2015}. Models that incorporated the next highest order exchange coupling $J_4$ yielded only marginal improvements to the fit quality as shown in Fig.~S6, motivating us to truncate the minimal model at $J_3$. Similarly, we do not consider dipolar interactions, which are two orders of magnitude weaker than our leading exchange term.

We fitted the spin wave dispersions of CrSb along all the high-symmetry directions of its BZ and obtained $J_1=23(4)~$meV, $J_2=-5.4(8)~$meV, $J_3=5.2(8)$~meV, and $D=0.15(4)~$meV. The spin wave spectrum calculated using our refined spin Hamiltonian is compared to the data in Fig.~\ref{fig:Magnon}(c,d) and shows good agreement along all reciprocal space directions. These values can be compared with previously reported parameters calculated using density functional theory (DFT)~\cite{PhysRevB.111.174451,kravchuk2025chiral}, as summarized in Table~SV.
Our experimentally determined exchange couplings differ appreciably from those of Ref.~\cite{kravchuk2025chiral} but show reasonable qualitative agreement with those of Ref.~\cite{PhysRevB.111.174451}. As shown in Fig.~S7, the parameters of  both Ref.~\cite{PhysRevB.111.174451} and Ref.~\cite{kravchuk2025chiral} reproduce the qualitative features but fail to capture the steepness of the dispersion and undershoot and overshoot, respectively, the total bandwidth by approximately 50~meV.

Having established a refined spin Hamiltonian that accurately reproduces the spin-wave dispersion along high-symmetry directions, we now turn to the more subtle effect of altermagnetism. In the case of the $g$-wave altermagnetism in CrSb, the $11^{th}$ and $12^{th}$ n.n. exchange pathways are equidistant (see Fig.~\ref{fig:Structure}(a)) but symmetry-inequivalent and can generate a chiral splitting of the magnon dispersion~\cite{PhysRevB.111.174451,kravchuk2025chiral,Beida2025}. Such a splitting is expected to occur along low-symmetry paths in the BZ, particularly along the $\Gamma$–$L$ direction, and becomes observable only when the difference between these two higher-order exchange couplings, $J_{11} - J_{12}$, is non-zero. Based on DFT calculations~\cite{PhysRevB.111.174451,kravchuk2025chiral}, we assumed $|J_{12}| > |J_{11}|$, and that $J_{12}$ is AFM.

To visualize this spin splitting, we simulated the spin-wave spectrum within the ($H$$+$$K$,$\text{-}2K$,2.85) scattering plane assuming $J_{11}=0$ and $J_{12}=0$, 1, 2, and 3~meV. These predictions are shown in Fig.~\ref{fig:Splitting}(a), alongside the experimental INS data acquired at an energy transfer of 180~meV in Fig.~\ref{fig:Splitting}(b). The emergence of six-fold spin-wave splitting is clearly established along the six symmetrically equivalent $(H, 0, 2.85)$ directions ($\Gamma$-BZ corner). In contrast, such splitting is absent along the $(K,\text{-}2K,2.85)$ directions ($\Gamma$-BZ edge center), where instead brighter, more localized intensity spots are observed. This six-fold distribution of spectral weight is a signature of $g$-wave altermagnets. It is important to note that the observed intensity modulation in this energy range cannot be produced by any exchange couplings below $J_{11}$ (Fig. S8 and S9)~\cite{PhysRevB.111.174451}.

Comparison between simulated spectra and experimental data (Fig.~\ref{fig:Splitting}(a,b)) reveals a clear six-fold symmetry in the spin excitations, in contrast to the symmetric spin-wave cones expected for $J_{12}=0$. This provides direct evidence for chiral spin splitting arising from altermagnetic interactions in CrSb. Taking 1D cuts along the splitting and non-splitting directions (Fig.~\ref{fig:Splitting}(c,d)), we can observe a clear two-peak structure in the former and a single maximum in the latter. To quantify the splitting, we fit the cut along the $(H,0,2.85)$ direction by two Gaussian peaks. The resulting best fit yields a separation of 0.30(2)~\AA$^{-1}$ (0.17(1) r.l.u.), with a peak FWHM of 0.23(2)~\AA$^{-1}$ (0.12(2)~r.l.u.), which can be compared to the instrumental resolution of 0.191(1)~\AA$^{-1}$. In contrast, a 1D cut along orthogonal ($K$,$-2K$,2.85) direction is described by a single Gaussian peak, confirming the directional nature of the splitting (Fig.~\ref{fig:Splitting}(d)). A SpinW fit of (Fig.~\ref{fig:Splitting}(b)) yields $J_{12}= 1.8(4)$~meV. Using this value, we calculated the expected magnon dispersion along the $\Gamma$–$L$ direction (Fig.~\ref{fig:Splitting}(e)) and compared it to the experimental data (Fig.~\ref{fig:Splitting}(f)). The energy splitting reaches 27(2) meV at $H = 0.20$ (SI Fig.~S10), which is comparable to the instrumental resolution (20 meV at the relevant energy, Fig.~S11) and therefore remains unresolved in the energy dimension.

It should be emphasized that the lack of a clear splitting signature in energy space is not due to the size of the splitting. In fact, the magnon splitting in CrSb is an order of magnitude larger than any previously measured altermagnet (27 meV for CrSb as compared to 3~meV in $\alpha$-Fe$_2$O$_3$~\cite{7yhz-jptc}) and is also the largest as a percentage of its total bandwidth (12\% as compared to 5\% for MnTe~\cite{liu2024chiral}) (see Table SVI). Rather, this measurement lies well outside the energy range over which flux and resolution are optimized in modern neutron spectrometers. The splitting detection method established here exploits the relatively better $Q$ resolution of such instruments and can be generally applied to any phenomena that yield lifted degeneracies in excitation spectra. Importantly, chiral magnon splitting is a symmetry-enforced property and therefore persists throughout the magnon spectrum, including the low-energy regime relevant for device applications. Beyond the evidence for splitting observed here, the chiral nature of CrSb's magnons has also been demonstrated through resonant inelastic x-ray scattering utilizing circularly polarized light~\cite{Biniskos2025}.

We report the first experimental detection of chiral magnon splitting in a metallic altermagnet, CrSb. Polarized neutron diffraction measurements confirm the fully compensated magnetic ordering below the N\'eel temperature. By mapping the magnon dispersion in CrSb, we uncover the microscopic exchange interactions that stabilize its magnetic order, enabling the determination of the spin Hamiltonian. The first three nearest-neighbor interactions well describe the main features of the spectrum, while the longer-range $11^{\text{th}}$ nearest-neighbor interaction is crucial for capturing the subtle altermagnetic signature. Altermagnetism in CrSb gives rise to coherent, spin-split magnons with large group velocity that persist well above room temperature. Together with its metallic character and weak damping, these properties provide a promising platform for spintronic applications, where efficient angular momentum transfer, directional magnonic responses, and fast switching dynamics are desirable.

\begin{acknowledgments}

The authors are grateful to Christine Trinh, Maxim Levi, and Issac Hoffman for their prior work on the crystal growth of CrSb. We also thank Cedric Gagnon for his help with the design of the sample mount used for the INS experiment. We thank Patrick Brian and Brodie Miles for their assistance in the SCXRD and SEM-EDS data collection, respectively. AKS was supported by the Killam Postdoctoral Fellowship. This work was supported by the Natural Sciences and Engineering Research Council of Canada (NSERC), the Canadian Institute for Advanced Research (CIFAR), the Sloan Research Fellowships program, and the Killam Accelerator Research Fellowship. This research was undertaken thanks in part to funding from the Canada First Research Excellence Fund, Quantum Materials and Future Technologies Program. A portion of this research used resources at the High Flux Isotope Reactor and Spallation Neutron Source, a DOE Office of Science User Facility operated by the Oak Ridge National Laboratory. The beam time was allocated to HB-1 and SEQUOIA on proposal numbers IPTS-34259 and IPTS-34137, respectively. The support for neutron scattering was provided by the Center for High-Resolution Neutron Scattering, a partnership between the National Institute of Standards and Technology and the National Science Foundation under Agreement No. DMR-2010792. The identification of any commercial product or trade name does not imply endorsement or recommendation by the National Institute of Standards and Technology. 
\end{acknowledgments}

\section*{Data Availability}

The raw polarized neutron diffraction data collected on HB-1 are openly available via the ORNL Neutron Catalog (ONCat) at \textcolor{blue}{https://doi.org/10.14461/oncat.data/17428543}. The raw inelastic neutron scattering data collected on SEQUOIA are openly available via ONCat at \textcolor{blue}{https://doi.org/10.14461/oncat.data/17428573}.

\bibliography{bibliography}

\end{document}